\documentclass[preprint,superscriptaddress,showpacs,amsmath,amssymb,aps,prx,floatfix]{revtex4-1}
\usepackage[utf8]{inputenc}
\usepackage{amsmath}
\usepackage{amsfonts}
\usepackage{amssymb}
\usepackage[super]{nth}

\usepackage{amsmath}
\usepackage{fourier}
\usepackage{color}
\usepackage{graphicx}
\usepackage{hyperref}% add hypertext capabilities
\usepackage{nameref}
\usepackage{soul}
\usepackage{setspace}

\begin{document}
\title{A sampling-guided unsupervised learning method to capture percolation in complex networks}

\newcommand{\RochesterP}{Department of Physics \& Astronomy, University of Rochester, Rochester, NY 14627, USA}
\newcommand{\RochesterC}{Department of Computer Science, University of Rochester, Rochester, NY 14627, USA}
\newcommand{\GG}[1]{{\color{blue} #1}}

\author{Sayat Mimar}                    
\affiliation{\RochesterP}
\author{Gourab Ghoshal}                    
\affiliation{\RochesterP}
\affiliation{\RochesterC}

\begin{abstract}
\singlespacing
The use of machine learning techniques in classical and quantum systems has led to novel techniques to classify ordered and disordered phases, as well as uncover transition points in critical phenomena. Efforts to extend these methods to dynamical processes in complex networks is a field of active research. Network-percolation, a measure of resilience and robustness to structural failures, as well as a proxy for spreading processes, has numerous applications in social, technological, and infrastructural systems. A particular challenge is to identify the existence of a percolation cluster in a network in the face of noisy data. Here, we consider bond-percolation, and introduce a sampling approach that leverages the core-periphery structure of such networks at a microscopic scale, using onion decomposition, a refined version of the $k-$core. By selecting subsets of nodes in a particular layer of the onion spectrum that follow similar trajectories in the percolation process, percolating phases can be distinguished from non-percolating ones through an unsupervised clustering method. Accuracy in the initial step is essential for extracting samples with information-rich content, that are subsequently used to predict the critical transition point through the confusion scheme, a recently introduced learning method. The method circumvents the difficulty of missing data or noisy measurements, as it allows for sampling nodes from both the core and periphery, as well as intermediate layers. We validate the effectiveness of our sampling strategy on a spectrum of synthetic network topologies, as well as on two real-word case studies: the integration time of the US domestic airport network, and the identification of the epidemic cluster of COVID-19 outbreaks in three major US states. The method proposed here allows for identifying phase transitions in empirical time-varying networks. 
\end{abstract}
	
	\maketitle

\section{Introduction}

Artificial intelligence has abundant applications in a wide spectrum of disciplines including health care, medicine, finance, autonomous driving and engineering of smart devices to name a few~\cite{jordan_machine_2015}. In the physical sciences, machine learning (ML) techniques have been used to extract useful information from massive datasets generated by particle physics experiments or observations in astronomy~\cite{mehta_high-bias_2019}. In condensed matter physics, ML methods have been adapted to study thermodynamic phase transitions in several classical systems such as the Ising Model \cite{Carrasquilla2017}, the $XY$ Model~\cite{Beach2018, Zhang2019} and the Hubbard model \cite{ChNg2018}, as well as to explore quantum phase transitions~\cite{Arsenault2014,Venderley2018,Che2020,Lidiak2020}. On regular lattices, for instance, unsupervised learning models are used to  discriminate between ferromagnetic and paramagnetic spin configurations at different temperatures, with unlabeled samples above and below the percolation threshold. Examples of such methods include principle component analysis (PCA)~\cite{Wang2016}, $t-$distributed stochastic neighboring ensemble ($t-$SNE)~\cite{Zhang2019}, $k-$means clustering ~\cite{Canabarro2019}, auto-encoders~\cite{Wetzel2017} and the recently introduced confusion scheme~\cite{VanNieuwenburg2017}. When labels with ordered-disordered states of configurations are introduced, supervised learning methods such as Artificial (ANN)~\cite{Carleo2017,morningstar} and Convolution Neural Networks (CNN)~\cite{Huembeli2018}, 
are employed to infer the transition temperature, which has been found to be in exact agreement with theoretical predictions \cite{KochJanusz2018,Chng2017}  
\par
In complex networks an important example of critical phenomena is percolation, a measure of structural resilience and a benchmark model for other dynamical processes such as epidemic spreading, vital node identification and community detection~\cite{Newman2002, derenyi_clique_2005, lu_vital_2016}. One example is bond-percolation, where in a network of $N$ nodes, $E$ edges are added randomly to an empty network (or conversely removed at random from a connected network), until at a critical fraction of edges $\phi_c$, a giant connected component (GCC) of size $O(N)$ emerges in a continuous second-order phase transition. While the critical bond-occupation probability $\phi_c$ can be computed using numerical~\cite{Newman2000} and analytical methods~\cite{Newman_2001, Newman_2008, Ghoshal_2009, Zlatic_2009}, for networks of smaller size and in real-world networks with incomplete data, these are less predictive compared to the true percolation threshold~\cite{Radicchi2015}.

%Besides, limited testing/tracing resources can impose challenges on obtaining the entire contact history of such influential spreaders \cite{Kong2021}. Recently, techniques have been proposed to study immunization strategies in networks with only a small subset of nodes are observed at a time to mitigate the effect of epidemic spreading \cite{Shang2021_1,Rosenblatt2020}. Under such circumstances of limited knowledge of network structure, immunizing a small sample of nodes provides significant improvement in the global level immunization of the network \cite{Liu2020}.

Indeed, data on empirical networks is commonly noisy and restricted to sub-samples~\cite{Otsuka2019}. Furthermore, uncertainty in measurements alters network topology and impacts structural measures such as centrality as well as the percolation threshold, which in turn affect network dynamics~\cite{Ghoshal_2011, martin_niemeyer_2019,NIU2015124, Mimar_2019, Mimar_2021}. In particular, nodes in the network-core are more sensitive to incomplete observations and missing links \cite{Platig2013,Shang2021}. Examples include Call data records (CDR's) from mobile-phones that miss connections due to missing phone numbers, as well as online social networks that may indicate existing virtual ties among people who are unacquainted~\cite{Onnela7332}. In the context of epidemics, super-spreading events have been identified as a significant source and driver of major outbreaks~\cite{Stein2011}. However, contact-tracing the network of spread is biased by limitations in  data collection and  public  health  capacity,  potentially leading  to over- or under-estimation of the extent of super-spreading \cite{Susswein2020}. 

Recently, ML techniques have been proposed to study the epidemic cluster in the susceptible-infectious-susceptible (SIS) compartmental model of epidemic spreading on networks \cite{Ni2019_2}. The proposed  approach converts high dimensional network data into image-like structures and exploits CNNs to learn and precisely identify the outbreak threshold of epidemic dynamics. Similar techniques were used for the case of  sparse time-series data of a handful of nodes produced by networks of coupled Kuramoto oscillators~\cite{Panday2021} to  accurately classify underlying network structure. In~\cite{Ni2019} a deep learning framework is introduced, combining both unsupervised and supervised learning methods to predict phase transitions associated with spreading dynamics. This approach makes accurate estimates of the critical transition point in uniform random networks, as well as proposes hub$-$and$-$neighbors and max$-k-$core sampling to overcome predictive inaccuracies associated with networks that have heavy-tailed distributions of links. Indeed, the rich structural features of heterogeneous networks render the simultaneous prediction of the critical transition point and the clustering of dynamical phases by adopting unsupervised learning approaches like PCA, challenging. Indeed, in complex hierarchical networks, separating percolating and non-percolating regimes in dynamical processes remains unsolved. 

%The method overcomes predictive inaccuracies in existing methods associated with those networks that have a heavy-tailed distributions of links. Yet,

To uncover the precise role of network structure in the learning process, in this manuscript, we focus on bond-percolation and investigate the effect of topology on ML methods that seek to estimate the percolation clusters and to infer the critical bond occupation probability $\phi_c$. Our approach extends previously proposed macro-level  sampling procedures by using onion decomposition (OD) ~\cite{Hebert-Dufresne2016} as a tool to determine the position of nodes in the core-periphery structure. This network statistic---a refined version of the $k-$core decomposition~\cite{Allard2019}---decomposes the network into hierarchically ordered layers and reveals much more structural information at meso and micro scales. We investigate the limiting cases of uniform--  and heavy-tailed--distribution and demonstrate significant  differences between networks of opposite topologies, with the former having a homogeneous population of nodes across layers, while the latter containing dense layers interspersed by sparse regions. We use a hybrid unsupervised learning method, combining $t-$SNE and $k-$means clustering, and train it on subsets of nodes sampled from both the sparse and dense layers to distinguish dynamical states above and below $\phi_c$. We show that sampling from the dense layers with nodes containing similar dynamical information in the percolation process, provides significantly higher accuracy than compared to sampling from the sparse layers or sampling nodes randomly, independent of whether nodes lie in the core or the periphery. 

\par
Having determined the optimal sampling strategy, we next use the confusion scheme to identify the critical occupation probability  $\phi_c$, and once again demonstrate high accuracy as compared to the ground-truth estimates of the threshold values. Perhaps, most surprisingly, we show that optimal samples are not limited to the core of the underlying network, but there exists multiple subset of nodes in the entire range of layers in the onion spectrum. This finding bears particular significance for using such methods in empirical networks, the majority of which have heavy-tailed distributions of links, and whose measurements are noisy. Finally, we apply our formalism to two examples of real-world examples: We determine the exact critical integration time of the US air transportation network , as well as identify the epidemic cluster for COVID-19 in three major states in the US. We end with a discussion of the implications of our findings.

\section{Percolation on different network topologies}
\begin{figure}[t]	
	\centering
	\includegraphics[width=1.\linewidth]{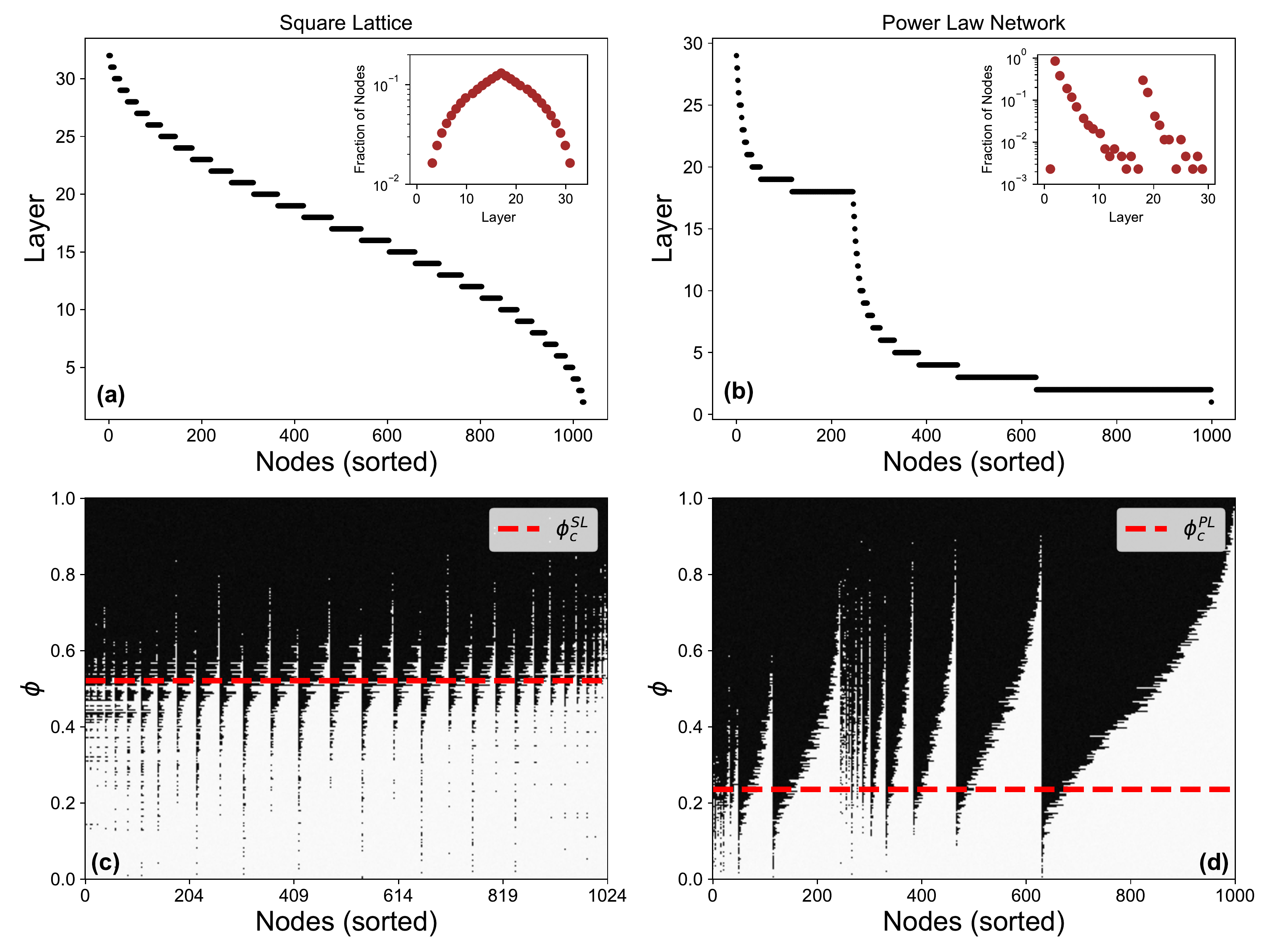}\\
	\caption{{\bf Effect of topology on percolation dynamics} Onion decomposition of \textbf{(a)} a Square Lattice of size $N = 1024~(32\times32)$ and \textbf{(b)} a power-law network with $N=1000$ and $\gamma = 3.1$. Nodes are ordered from the inner- to the outer-most layers which are labeled in descending order. The onion spectrum is shown as inset, indicating that the square-lattice consists of a single shell, whereas the power-law network contains three distinct shells. Layers are populated uniformly in the square-lattice, and in a punctuated fashion in the power-law network. In \textbf{(c)} and \textbf{(d)}, the percolation process in the range $0 \leq \phi \leq 1$ for the square-lattice and power-law network. The critical bond-occupation probabilities $\phi_c^{SL} = 0.524$ and $\phi_c^{PL} = 0.248$ are marked as the red dashed line. Nodes part of the GCC are colored black, those outside are colored white. Nodes are ordered the same as in the upper-panel.}
	\label{fig:data_onion}
\end{figure}
Consider a network $G$ where $\mathcal{V}=\left\{v_{1}, \cdots, v_{N}\right\}$ is the node set that undergoes bond-percolation, and let time $t$ denote the dynamical state of a certain configuration of occupied bonds. The data $\textbf{X}$ generated during the process is contained in a $M \times N$ matrix, where $N$ is the number of nodes and $M$ is the total number of dynamical states at different values of the occupation probability $\phi$. The entries of $\textbf{X}$ are binary;
\begin{equation}
x_{(v, t)}=\left\{\begin{array}{ll}
1, & v \in GCC \\
0, &v \notin GCC.
\end{array}\right.
\label{eq:input}
\end{equation} 
Each value is characterized by the tuple $(v, t)$ and equals $1$ if $v$ is part of the giant connected component (GCC) of the network and $0$  if the node is disconnected from GCC at time $t$ where $1\leq t \leq M$. We pick two graphs at the opposite ends of the structural spectrum of networks: first, we construct a square lattice with open boundaries such that all nodes, except those at the periphery, have degree $k=4$, leading to a uniform, tightly-peaked degree distribution. On the other end of the spectrum, we consider a power-law network with degree distribution $p_k \sim k^{-\gamma}$ generated using the configuration model~\cite{molloy_critical_1995}. We choose an exponent $\gamma = 3.1$, such that both networks have finite second moments $\langle k^2 \rangle$ in their degree-distributions and exhibit phase transitions at non-zero probabilities $\phi_c = \left[\langle k^2 \rangle/\langle k \rangle -1\right]^{-1}$ in the thermodynamic limit~\cite{Newman_2001}.
\par
Next, we use the onion decomposition method to uncover the core-periphery structure of the two different types of networks. In addition to the \textit{coreness} metric produced by $k$-core decomposition that identifies nested maximal subnetworks with nodes having at least $k$ connections, the onion decomposition improves the \textit{coreness} information by assigning a layer to each node, to further indicate its position within the core and make its internal organization apparent~\cite{Hebert-Dufresne2016}. 

In Fig.~\ref{fig:data_onion}, panels $\textbf{a}$ and $\textbf{b}$ we show the onion spectra of the square lattice and the power-law network. Nodes are sorted with respect to their layer value in descending order, from the inner- to the outer-most layer. The square lattice has a uniform spectrum where nodes populate each layer equally in the network, however the power-law network shows a spectrum with sparse-filling in some of the inner and middle layers, punctuated by dense intermediate and peripheral layers. The fraction of nodes in each layer (shown as inset in both panels) indicates that the square-lattice contains all nodes in a single-shell, whereas the power-law network contains at least three distinct shells. The effect of this difference in structure is shown in panels {\bf c} and {\bf d}, where we show the evolution of the percolation process in the range $0 \leq \phi \leq 1$. The critical bond occupation probabilities $\phi_c^{SL}$ and $\phi_c^{PL}$, below which there is no GCC, are marked by the red horizontal dashed lines. The horizontal axis corresponds to the nodes sorted in the same order as in the upper panel. Nodes that are part of the GCC are colored black, whereas those outside the GCC are colored white. (Note that the few sets of nodes colored black below $\phi_c$ belong to the largest connected component (LCC) which is technically not the GCC.)
\par
The figure indicates that in the case of the square lattice, groups of nodes across layers show common dynamics for a wide-range of $\phi$. Nodes attach and detach from the GCC (as $\phi$ changes) in a similar fashion independent of what layer they belong to. In contrast, in the power-law network there is wide variation across layers; for a given value of $\phi$ large swathes of nodes in the inner- and outer-most layers are not part of the GCC. Further, whether a node is part of the GCC as $\phi$ is increased, varies from layer-to-layer. Finally, we note the presence of high-fidelity samples in the core-, intermediate- and peripheral-layers. For any classification algorithm, such class-imbalanced datasets pose a challenge as learning methods fail to capture the distributive characteristics of the data and produces unsatisfactory accuracies~\cite{HaiboHe2009}. The performance of such algorithms is poor on subsets with under and over-represented classes as it tends to partition phases into relatively uniform sizes. The pre-processing of the data using the onion decomposition method, instead allows for the identification of of node subsets with similar dynamical information in the percolation process. That is, layers where nodes disconnect from the GCC at comparable values of the control parameter, hence yielding a balanced training data in the subsequent learning phase.

\section{Effect of sampling on clustering}
\label{sec:clustering}

Next, we focus on the classification scheme for clustering nodes as part of- or excluded from- the GCC.  We construct a hybrid unsupervised learning model with $t-$SNE \cite{JMLR:v15:vandermaaten14a} a non-linear dimensionality reduction technique, and $k-$means clustering~\cite{kriegel_densitybased_2011} used for identifying pre-determined number of clusters from an unlabeled dataset. We add Gaussian noise, $\mathcal{N}(0,\,0.01)$, to the input data $\textbf{X}$ (Eqn.~\ref{eq:input}) to help spread the data points and sample subsets of nodes in bins of size $20$ (corresponding to $\approx 50$ samples), ranging from the innermost layer to the peripheral layer. We project the subset of 20 nodes into a two-dimensional plane with $t-$SNE, and then use $k-$means with $k = 2$ to assign labels as in Eq.~\eqref{eq:input}. To assess the performance of the algorithm we compare the labels assigned by the unsupervised learning method $\hat y$ to the ground-truth label $y$ and define the accuracy $\alpha_{\hat y, y}$ as 
\begin{equation}
\alpha_{\hat y, y}=\frac{1}{n_{\text {samples }}} \sum_{i=1}^{n_{\text {samples }}} \delta_{\hat y, y},
\label{eq:accuracy}
\end{equation}
where the summand is the Kronecker delta-function. 
\par
\begin{figure}[t!]	
	\centering
	\includegraphics[width=1.\linewidth]{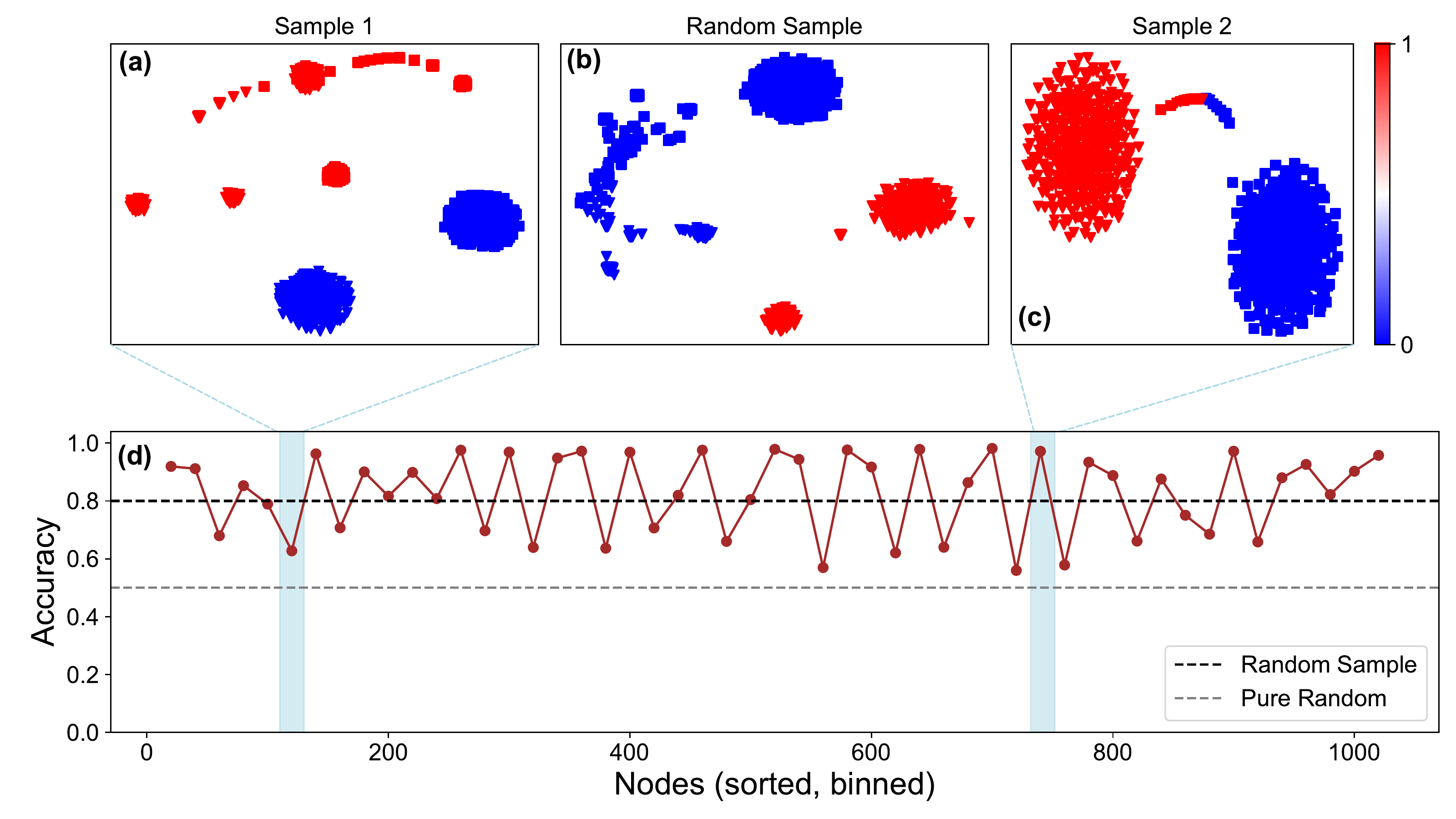}\\
	\caption{{\bf Effect of sampling on the unsupervised learning scheme for the square lattice.} \textbf {(d)} The accuracy of label prediction $\alpha_{\hat y, y}$ with $t-$SNE and $k-$ means clustering as a function of sampling layers, binned in sets of 20 nodes, and ordered the same as Fig.~\ref{fig:data_onion}. The accuracy for random samples of 20 nodes ($\alpha_{\hat y, y} = 0.8$) is shown as a black dashed line, and the grey dashed line corresponds to model-independent random guessing of state labels ($\alpha_{\hat y, y} = 0.5$). Examples of clustering \textbf {(a)} for low-accuracy samples from the inner-layer \textbf{(b)} random samples and \textbf{(c)} high-accuracy samples from the outer-layer. }
	\label{fig:unsup_lat}
\end{figure}

In Fig.~\ref{fig:unsup_lat} we plot the results of our analysis for the square-lattice. Panel \textbf{a} shows $\alpha_{\hat y, y}$ as a function of the sampled subset of nodes, ordered the same as in Fig.~\ref{fig:data_onion}. The horizontal black dashed line indicates the accuracy for random samples of 20 nodes ($\alpha_{\hat y, y} = 0.80$), and the gray dashed line represents the accuracy for a model-independent random guess of the state-labels ($\alpha_{\hat y, y} =0.5$). As the figure indicates, depending on the sampled layer, the accuracy fluctuates around the performance of the random sampling method, with some layers providing almost perfect accuracy, whereas others no better than a random guess of labels. The high-and low-accuracy samples are not limited to the core, but periodically found across all layers in the onion spectrum. Example outputs of the unsupervised learning scheme are shown in panel {\bf a} (inner-layer, $\alpha_{\hat y, y} =0.56$), {\bf b} (random sample, $\alpha_{\hat y, y} =0.80$) and {\bf c} (outer-layer, $\alpha_{\hat y, y} =0.99$). For the poor-accuracy inner-layer sample, the model is confused by the fact that nodes in the training-set exhibit different dynamical evolution in the percolation process. They connect to the GCC at different values of $\phi$, hence the unsupervised learning model cannot cluster the configurations into two separate clusters. Conversely, for the high-accuracy outer-layer sample, the subset of nodes belong to a high-fidelity layer in the onion spectrum, such that the majority of nodes in the set, connect to the GCC at similar values of $\phi$. Such layers, as indicated in Fig.~\ref{fig:data_onion}{\bf c} are distributed equally across the onion spectrum. Finally, the random sampling strategy provides reasonable accuracy, given the homogenous structure of the square lattice (all nodes in a single-shell), as random samples and tailored subsets have similar properties. 

\begin{figure}[t!]	
	\centering
	\includegraphics[width=1.\linewidth]{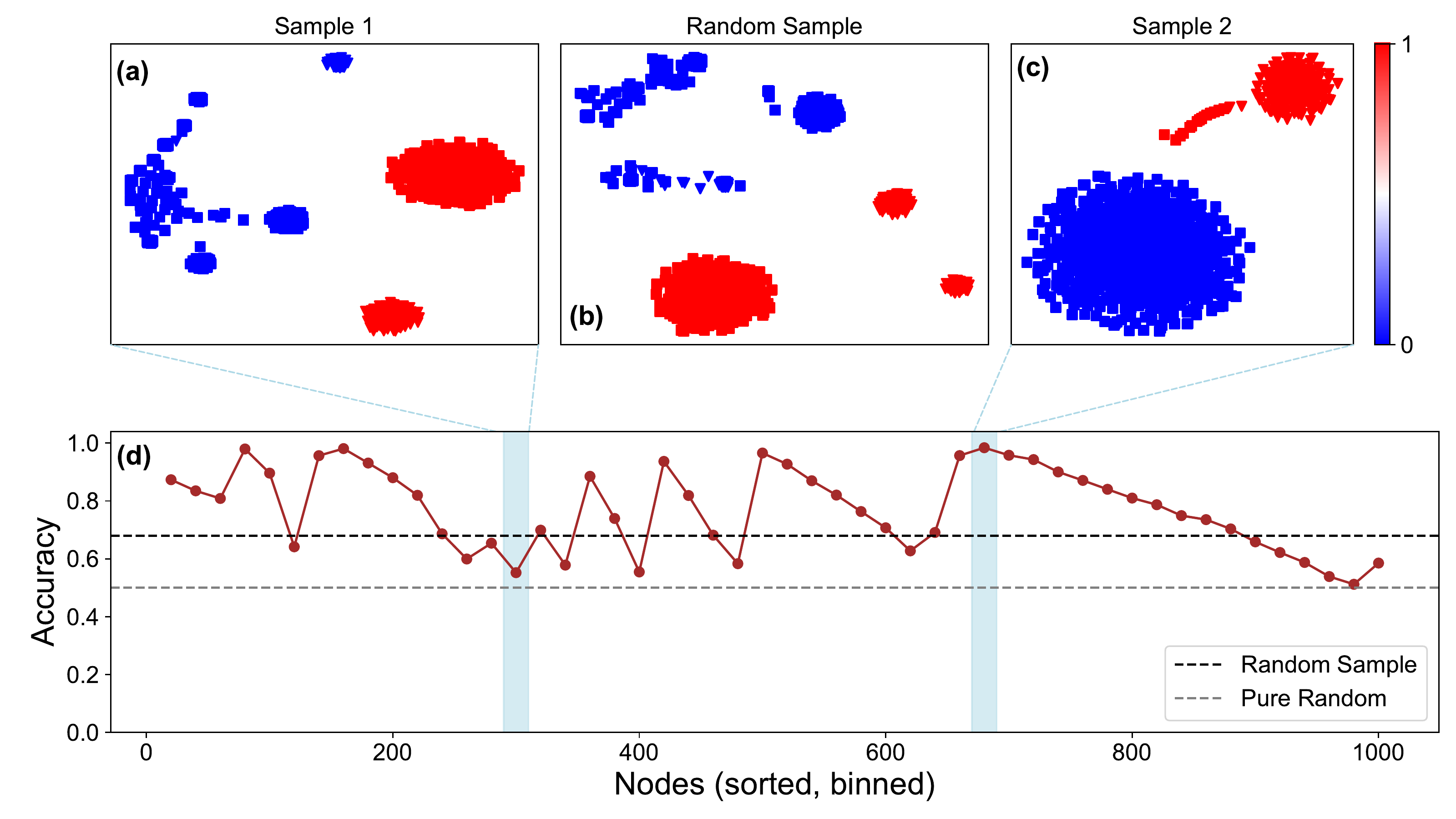}\\
	\caption{{\bf Effect of sampling on the unsupervised learning scheme for the power-law network} \textbf {(d)} The accuracy of label prediction $\alpha_{\hat y, y}$ with $t-$SNE and $k-$ means clustering as a function of sampling layers, binned in sets of 20 nodes, and ordered the same as Fig.~\ref{fig:data_onion}. The accuracy for random samples of 20 nodes ($\alpha_{\hat y, y} = 0.64$) is shown as a black dashed line, and the grey dashed line corresponds to model-independent random guessing of state labels ($\alpha_{\hat y, y} = 0.5$). Examples of clustering \textbf {(a)} for low-accuracy samples from the inner-layer \textbf{(b)} random samples and \textbf{(c)} high-accuracy samples from the outer-layer. }
	\label{fig:unsup_sf}
\end{figure}

In Fig.~\ref{fig:unsup_sf} we plot the corresponding results for the power-law network indicating rather different behavior. As seen in panel {\bf b}, the random sampling strategy performs considerably poorer with $\alpha_{\hat y, y} = 0 .64$, only marginally better than randomly guessing labels. Furthermore, the peaks and troughs in the accuracy curve are much more irregular, as compared to the square-lattice, reflecting the richer structure of the power-law network. Wide ranges in the intermediate-layer provide poor accuracy (panel {\bf a}), and surprisingly, there exists a range in the outer-most layers that yields accuracy as high as 0.99 (panel {\bf d}). The results indicate the importance of adopting a considered sampling strategy for training-sets in power-law topologies, given that unlike in networks with uniform topologies, random sampling is sub-optimal. Indeed, very few real-world networks have uniform topologies, instead exhibiting heavy-tailed distributions, implying that for any realistic application, identifying high-quality samples \emph{a priori} is of paramount importance. Given issues of data sparsity, it is of note, that such samples exist in multiple layers of the power-law network, including the core-, intermediate- and peripheral layers.

\par

%Illustrations of the unsupervised clustering method at certain subsets are shown in panels \textbf{b}, \textbf{c}, \textbf{d} of Fig.~\ref{unsup_lat} and Fig.~\ref{unsup_sf}. In panel \textbf{b}, result on a subset with poor accuracy of $0.56$ and $0.58$ for square lattice and scale-free network respectively. In both figures, the model is confused by the fact that nodes in the training data have vastly different dynamical evolution in the percolation process. They disconnect from the GCC at distinct $\phi$ values, hence the unsupervised learning model cannot cluster the configurations into 2 separate clusters. Accuracy improves for square lattice when the training data is replaced by random samples, but clear clustering is still absent for scale-free network. This is due to the uniform structure of the square lattice as random samples and particularly chosen subsets have similar properties, thus the model is able to label snapshots with an accuracy of $0.80$ as shown in Fig.~\ref{unsup_lat} panel \textbf{c}. This is not the case for the scale free network which results in low accuracy of $0.64$ on the random samples illustrated in  Fig.~\ref{unsup_sf} panel \textbf{c}. Finally, with samples chosen from the same layer in consecutive order, the nodes have very similar percolation trajectories. In panels \textbf{d} of both figures, we achieve perfect clustering with accuracy $0.90$ for square lattice and $0.99$ for scale-free network.

\begin{figure*}[t]	
	\centering
	\includegraphics[width=1.\linewidth]{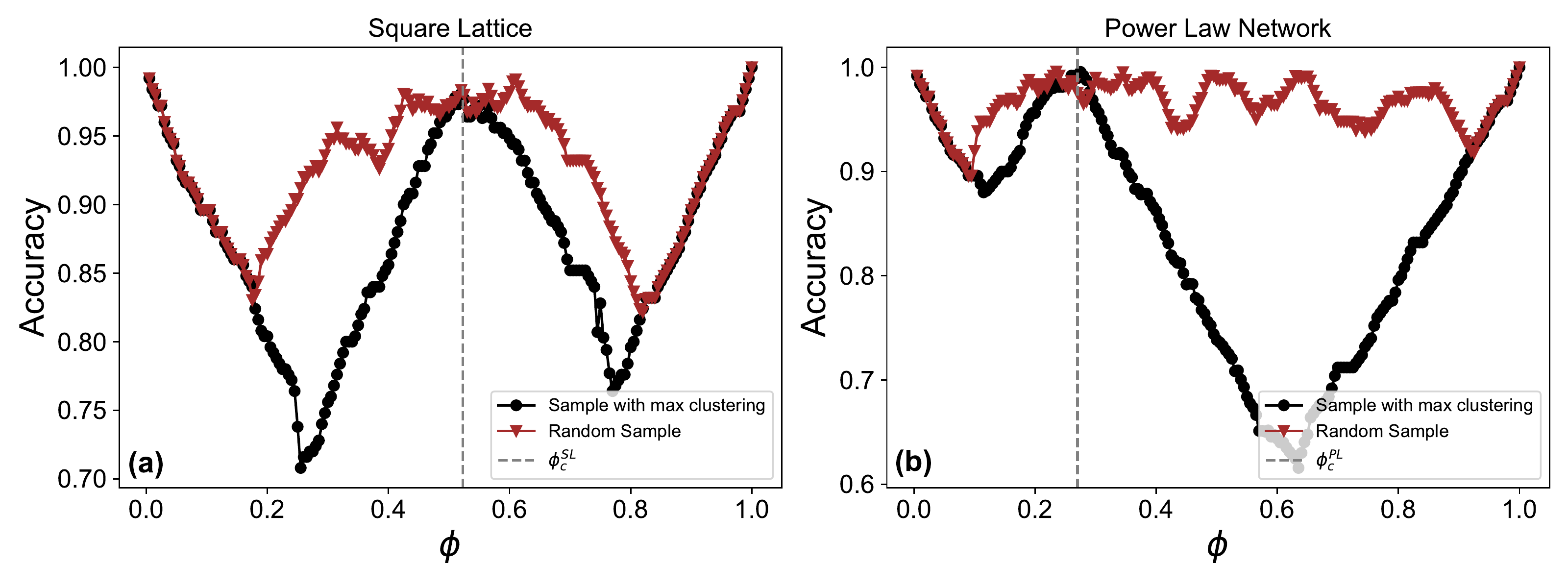}\\
	\caption{{\bf Identification of  $\phi_c$ as a function of network topology:} \textbf{a} Square lattice and  \textbf{b} power-law network with random samples of 20 nodes (brown curve) and a high-accuracy sample from the peripheral layers identified using the unsupervised learning method (black curve). The vertical dashed line represents the critical bond occupation probabilities in each network, $\phi_c^{SL} = 0.524$ and $\phi_c^{PL} = 0.248$. The random sample in the square lattice has a clear \textbf{W} shape, but the middle peak is noisy and is flat around $\phi_c^{SL}$. For the power-law network the random sampling strategy fails to provide any information on the transition probability. In both cases, the sampling-guided strategy yields high accuracy on $\phi_c^{SL,PL}$, with the middle peaks in the black curves occurring at 0.520 and 0.240.}
	\label{confusion_sf_lat}
\end{figure*}
\section{Identifying Critical Transition Points}

The procedure described thus far, while effective in identifying configurations below and above the percolation phase, in itself, cannot identify the critical bond occupation probability $\phi_c$. To do so, we make use of the confusion scheme, first introduced to study phase transitions in Kitaev chains, the classical Ising model and in disordered quantum spin chains~\cite{VanNieuwenburg2017}. Recently it has been extended to uncover the critical transition probability in dynamical phase transitions in complex networks~\cite{Ni2019}. Next, we show that our sampling-guided strategy adopted to the confusion scheme is quite effective in terms of identifying the value of $\phi_c$. 

The method does not take as input labels of the dynamical states, instead a synthetic label space $\textbf{y}$ is associated with an input matrix $\textbf{I}$, with entires 0 and 1, corresponding to the before and after states in the percolation process.  At $\phi = 1$ the label vector $\textbf{y} = \textbf{0}$ i.e. all configurations are in the before state, and for $\phi = 0$, the vector $\textbf{y} = \textbf{1}$, each snapshot is labeled as after state. The boundary between 0 and 1 (corresponding to the critical threshold) in this artificial label-set is varied  in the entire range of the control parameter $\phi \in [0,1]$ with increments of $\Delta = 0.005$ (yielding $200$ steps in total) and associated with a optimal subsample $\textbf{I} \subset \textbf{X}$ selected using the method described in Sec.~\ref{sec:clustering}. For both the square-lattice and the power-law network we select samples from the periphery that yield high accuracy. 

A feed-forward neural network (FFNN) is  trained with this data consisting of pairs $\{\textbf{I} : \textbf{y} \}$ in the form of supervised learning problem using the PyTorch library~\cite{NEURIPS2019_9015}. The input layer contains neurons at the same number of the chosen subsample size, followed by a hidden layer of 128 neurons. Both layers have rectified linear unit (ReLu) as activation functions. The output layer contains a single neuron with sigmoid activation function, that predicts the probability of a configuration belonging to one of the states. A binary cross-entropy loss-function is minimized in training, which is well suited for binary classification problems. For stochastic optimization we us the Adam method \cite{Kingma2015} with learning rate $10^{-3}$. To prevent over-fitting we use Dropout regularization with probability $10^{-1}$. In each step, the dataset is split into a training set that is fed into
the FFNN and prediction accuracy is evaluated on the test set. Highest accuracies are achieved at endpoints of the threshold range due to the constant nature of the label space. As one spans the range between $[0,1]$, initially lower accuracy values are observed as some configurations are associated with incorrect labels. At the transition probability, the artificial label space matches the ground truth, leading to a high classification accuracy. In this process, the accuracy curve follows a \textbf{W}-shape as a function of $\phi$, where the middle peak corresponds to the estimated transition probability~\cite{Ni2019} . 

\par
In Fig.~\ref{confusion_sf_lat}, we show the output of the confusion scheme on the square lattice \textbf{(a)} and the power-law network \textbf{(b)}. In both panels, the brown curve corresponds to a random sample of 20 nodes and the black curve to the high-accuracy subset from the peripheral layers, identified using the unsupervised learning method.  The vertical dashed lines mark the value for the ground-truth value of $\phi_c$ in each network. The random sampling strategy in the square lattice is reasonably effective, generating a ${\bf W}$-shape, although the peak near $\phi_c$ is not well-defined. In the power-law network the random sampling accuracy curve is noisy and flat yielding little-to-no information on the transition probabilities. However, in both cases the black curve yields a clear ${\bf W}$-shape and the middle-peaks line up well with the ground-truth values of $\phi_c$. Thus unlike existing methods, the sampling-guided scheme outlined here simultaneously identifies nodes in the GCC as well as provides accurate estimates for the bond-occupation probability. 

\section{Transitions in time-varying real-world networks}

\begin{figure}[t]	
	\centering
	\includegraphics[width=1.\linewidth]{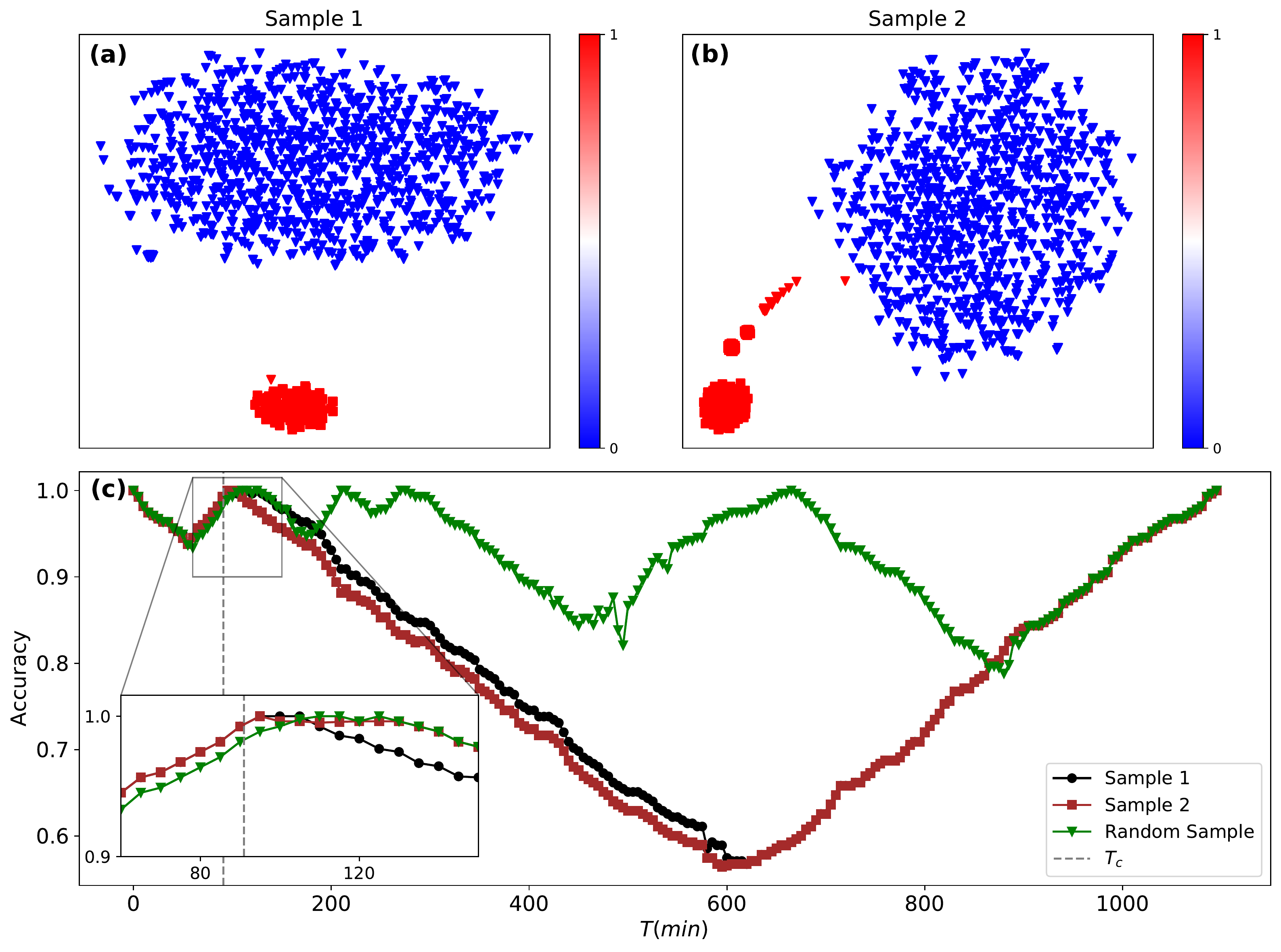}\\
	\caption{{ \bf Predicting integration time of the Unites States air-transportation network.} Clustering of pre- and post-transition states in the temporal integration process from a sample taken from the inner-layer (Sample 1) \textbf{a} and one from the peripheral layer (Sample 2) \textbf{b}. In both cases the accuracy $\alpha_{\hat y, y} = 0.99$. In \textbf{c}, the output of the confusion scheme in predicting the critical integration time ($T_c = 91$ mins, marked as vertical dashed line) at which a connected cluster of airports linked by flights is formed. Training the network on Sample 1 yields a peak of the \textbf{W}-shaped curve (black-circles) at $T = 93$ mins and with Sample 2 (red squares), at $T = 94$ mins. Training on a random sample (green triangles) yields a noisy curve, and an accurate identification of $T_c$ is not possible. (Behavior of all three curves near $T_c$ shown as zoomed inset.)}
	\label{air_res}
\end{figure}

Next, we validate our methodology in two real-world examples. The availability of large time-resolved datasets enables for the representation of a wide range of dynamical phenomena in the form of time-varying networks. Such processes often exhibit phase transitions, and thus can be analyzed via percolation in static networks. Some applications include wireless communication networks with unreliable links \cite{Karschau2018}, spreading of infectious diseases in modular time varying networks \cite{Nadini2018} or percolation in ground-transportation networks to identify critical bottleneck roads in local flows \cite{Li2015, Kirkley_2018}. Recently, the integration process of air traffic into a temporally connected network was modeled as as a time-varying percolation process~\cite{LIU2020_1}. The critical integration time $T_c$, at which the network forms a temporal spanning cluster, is proposed as a measure of global reliability of air-traffic. 

\par
We test our scheme on the air-transportation network, to identify both $T_c$ as well as label nodes that belong to the time-varying GCC. We construct the temporal air-transportation network \cite{air-data} starting from $t_0 = 7:30 AM$ on September 5, 2019 and spanning a 18-hour period until the integration process is completed. The resulting network consists of 288 airports as nodes and 1903 edges, where a link corresponds to at least one flight between two airports. We generate states with a time window of $ \left[t_{0}, t_{0}+T\right]$ where $T$ is  incremented in intervals of $1$-minute generating $\sim$1100 instances as the training and test set.
\par
The task of the unsupervised learning method is to discriminate between states before and after the integration process. Unlike in the synthetic networks studied thus far, the transition happens at an early stage ($T_c = 91$ mins), and therefore the dataset is unbalanced. After running the onion-decomposition scheme to identify the layers, we pick two samples with 10 airports: Sample 1 from the core of  the air transportation network \emph{(Atlanta, Austin, Nashville, Boston, Charlotte, Denver, Detroit, Fort Lauderdale-Hollywood, Las Vegas, Los Angeles, Chicago O'Hare)} and Sample 2 from the periphery \emph{(Norfolk, Worcester, Southwest Oregon, Barkley, Palm Beach, Hilton Head, Punta Gorda, Pitt-Greenville, Newport News/Williamsburg,  Ithaca Tompkins)}. We then use the $t-$SNE method to cluster the states. The results are shown in Fig.~\ref{air_res} panels \textbf{a} and \textbf{b}, indicating that the method performs well; the labels are assigned by $k-$means with an accuracy of $\alpha_{\hat y,y} = 0.99$ in both samples of nodes. We then use these two samples as a training set on the confusion scheme, and plot the resulting accuracy curve as a function of the time $t$ in Fig.~\ref{air_res}\textbf{c}. The curve corresponding to the core-sample is shown as black circles, while the that for the peripheral sample is shown as red-squares. As a reference we show the case for a random sample of 10 airports \emph{(Reno-Tahoe, Albany, Rapid City Regional, Hector, Miami, Waterloo, Kansas City, Corpus Christi, Columbia Metropolitan, Baltimore/Washington)} as green triangles, whereas $T_c$ is shown as the vertical dashed line. For both samples, the accuracy curve $\alpha_{\hat y, y}$ shows a clear ${\bf W}$-shape with peaks at $T= 93$ mins and $T= 94$ mins. The random sample yields a noisy curve with no clear peak (behavior near peak for all three curves shown as inset). The results illustrate the versatility of the sampling-scheme with near-perfect identification of airports in the temporally connected cluster and the ability to identify the critical integration time to within $3-4\%$, and the flexibility in sampling from the core or the periphery of the network.  
\begin{figure}[t]	
	\centering
	\includegraphics[width=1.\linewidth]{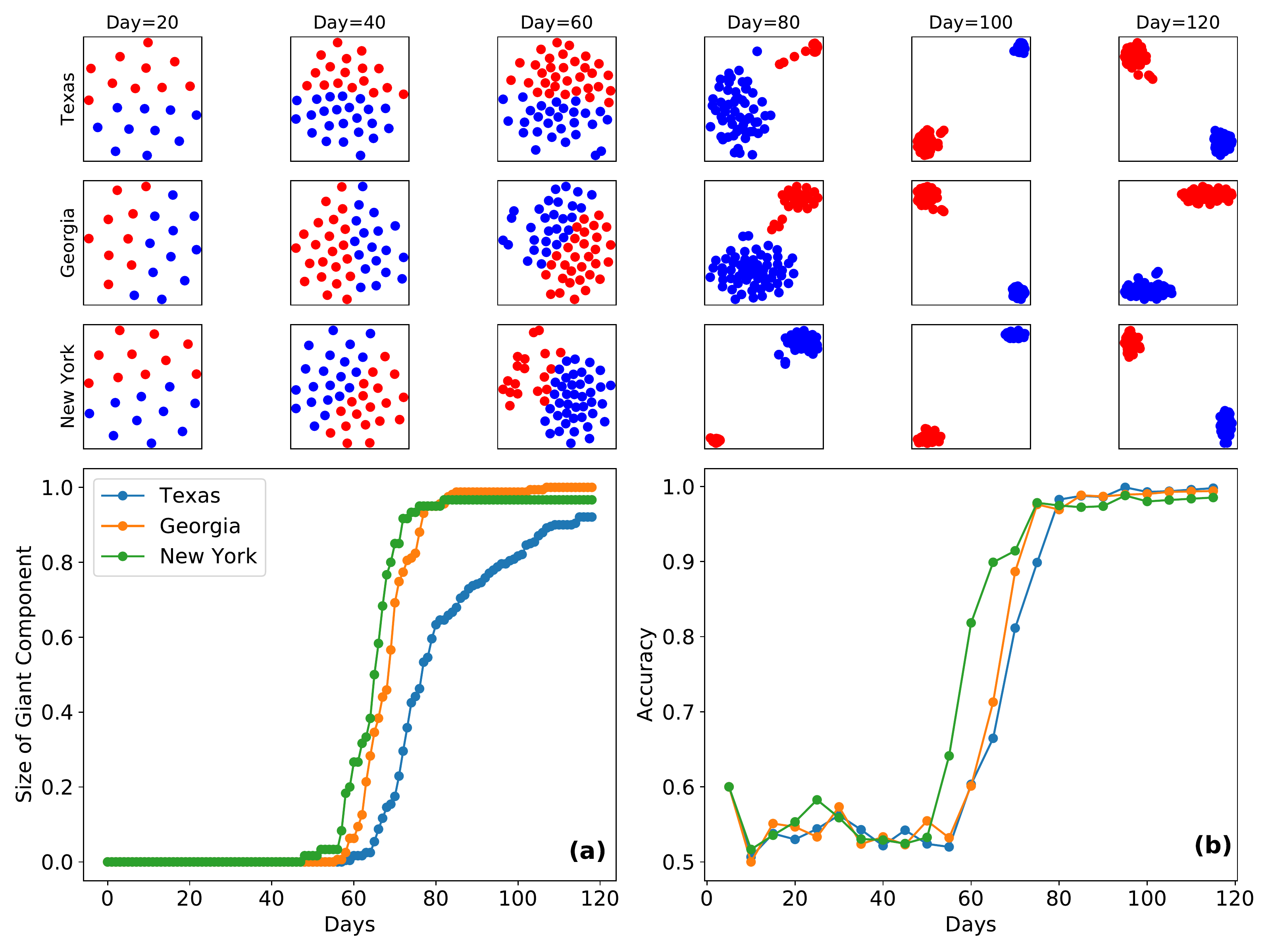}\\
	\caption{{\bf Emergence of the epidemic cluster of Covid-19 in 3 major US states.} The network corresponds to population flows between locations in-state at the resolution of counties. Nodes are labeled as infected if the number of cases per-capita exceeds $10^{-4}$. Panel \textbf{a} indicates that a epidemic cluster emerges around day 60 in all three states. The top three rows show the output of the clustering method that splits counties into those belonging to the epidemic cluster and those outside. Training occurs from day 0 in cumulative intervals of 20.  The identification of counties belonging to the epidemic cluster becomes increasingly accurate past the phase-transition and peaks around day 80 as seen in panel  \textbf{b}.}
	\label{covid_fig}
\end{figure}

\par 
Next, we consider a different dynamical process of particular relevance; the spread of COVID-19 in the United States~\cite{aguilar2020impact}. We investigate the possibility of employing our clustering method to use as a diagnostic tool that signals at a relatively early stage, whether an epidemic outbreak is about to occur based on real-time data. We pick three major states; Texas, Georgia and New York with 3.0 , 1.1 and 1.9 ($\times 10^7$) inhabitants respectively.  We consider a spatial resolution at the level of counties leading to 254 nodes for Texas, 159 for Georgia and 62 for New York. We construct mobility networks from the United States census bureau's LODES \cite{mobility-data} commuting data, where the edges represent population-flows between counties corresponding to  20262 (Texas), 11042 (Georgia) and 1883 (New York) undirected links. We then follow the temporal evolution of the number of detected cases in each county from January \nth{21} to May  \nth{18} 2020 \cite{covid-data}. Counties are labeled ``infected" when the number of cases per-capita is above a threshold of $10^{-4}$.  In Fig.~\ref{covid_fig}\textbf{a}, we plot the empirical temporal evolution of the number of infected counties finding an emergence of an epidemic cluster around day 60 for all three states. In the top-row of Fig.~\ref{covid_fig}, we train our unsupervised learning model with samples of size 10 selected from the high-fidelity layers of of the mobility networks in cumulative intervals of 20 days starting from day 0.  In Fig.~\ref{covid_fig}{\bf b} we plot the accompanying accuracy curve $\alpha_{\hat y, y}$ in function of time. Before day 60, there is no epidemic cluster, and therefor $\alpha_{\hat y, y} = 0.5$ equivalent to model-independent simple guessing of labels. After day 60, however, once an epidemic cluster emerges, the model is able to reliably split counties into infected and disease-free clusters and make more accurate predictions. The increase in the accuracy curve tracks the increase in the epidemic cluster and reaches perfect accuracy at around day 80. We note that from a point of real-time forecasting the model reaches accuracies of $\approx 70\%$ when the size of the epidemic cluster is $\approx 0.4$.

\section{Discussion}
Taken together, our work sheds light on the role of the micro- and mesoscopic structure of networks in machine learning the phases in bond percolation. Our sampling guided approach reveals the importance of choosing particular subset of nodes from layers of the onion spectrum of the network that enables unsupervised learning methods  to distinguish between percolating and non-percolating states. Labels assigned by $k-$means matches the ground truth labels with near-perfect accuracy. We show that this is facilitated by sampling nodes chosen from both core and peripheral layers, identified using onion decomposition, that identifies subsets of nodes in a spectrum of layers, that follow similar paths in the percolation process, i.e. they detach and attach to the percolating cluster at comparable values of $\phi$. The sampling-guided strategy carries over to other learning tasks, such as identifying the critical occupation probability $\phi_c$ using the confusion scheme. This gain in performance is particularly pronounced for networks with heavy-tailed degree distributions, where the method significantly outperforms random sampling. Indeed, to the best of our knowledge, the framework presented here is the first to simultaneously enable the clustering of nodes into different dynamical states, as well as identify $\phi_c$ in networks with heterogenous topologies. This bears significance, given that many empirical networks exhibit right-skewed distributions of links. 

\par
To validate our results, we demonstrate two possible applications of our findings on real-world time-varying networks that exhibit a percolation transition; the exact integration time of the US domestic air transportation network, as well as the emergence of the COVID-19 epidemic cluster in three large US states. In both cases the framework yields excellent performance. The application to pandemic settings is of particular interest, as a possible diagnostic tool to assess the current state of disease-spread with real-time data. The ability to accurately classify (with reasonable accuracy) regions into infected and disease-free states (close to when the epidemic cluster first emerges) could prove useful in terms of mitigation strategies. Indeed, techniques have been proposed to study immunization strategies in networks where only a small subset of nodes are observed at a time, to slow-down epidemic spread~\cite{Shang2021_1,Rosenblatt2020}. Given the limited knowledge of network structure, immunizing a small sample of nodes provides significant improvement in the global level immunization of the network \cite{Liu2020}. Similar considerations apply in the diffusion of rumors or ``fake news" in social media and online platforms~\cite{Lazer_2018}. The approach proposed here, can in principle be easily extended such types of dynamical processes on networks.

\bibliographystyle{naturemag}
\bibliography{perc_ref_v5}
	
\end{document}